\documentclass[prl,twocolumn,showpacs,preprintnumbers,amsmath,amssymb]{revtex4}
\usepackage{graphicx}
\usepackage{dcolumn}
\usepackage{bm}
\def\Term#1 #2 #3/{\mbox{$\,^{#1}\!#2_{#3}$ }}
\def\Termo#1 #2 #3/{\mbox{$\,^{#1}\!#2^o_{#3}$ }}

\begin{document}
\title{Ultrastable Optical Clock with Neutral Atoms in an Engineered Light Shift Trap}

\author{Hidetoshi Katori}
\altaffiliation[Corresponding author:]{ katori@amo.t.u-tokyo.ac.jp}
\author{Masao Takamoto}%
\affiliation{%
Engineering Research Institute, The University of Tokyo, Bunkyo-ku, Tokyo 113-8656, Japan}%

\author{V.~G.~Pal'chikov}
\affiliation{Institute of Metrology for Time and Space at
National Research Institute for Physical-Technical and
Radiotechnical Measurements -IMVP VNIIFTRI, Mendeleevo, Moscow
Region, 141579 Russia
}%
\author{V.~D.~Ovsiannikov}
\affiliation{Department of Physics, Voronezh State University,
Voronezh 394006, Russia
}%
\date{\today}
\begin{abstract}
An ultrastable optical clock based on neutral atoms trapped in an optical lattice is proposed.
Complete control over the light shift is achieved by employing the $5s^2 \,{}^1\!S_0 \rightarrow 5s5p \,{}^3\!P_0$ transition of ${}^{87}{\rm Sr}$ atoms as a ``clock transition".
Calculations of ac multipole polarizabilities and dipole
hyperpolarizabilities for the clock transition indicate that the contribution of the higher-order light shifts can be reduced to less than 1 mHz, allowing for a projected accuracy of better than $ 10^{-17}$.
\end{abstract}
\maketitle
Careful elimination of perturbations on electronic states and of motional effects has been considered as a prerequisite for realizing an atom frequency standard \cite{Luiten}.
A single ion trapped in an RF quadrupole field is one of the ideal systems that satisfy these requirements \cite{Dehmelt}, as the trap prepares a quantum absorber completely at rest in free space for an extended time and its electric field vanishes at the center of the trap.
Employing this scheme, quantum projection noise (QPN) limited spectroscopy \cite{Wineland1992} has been performed with an expected accuracy of $10^{-18}$ \cite{Luiten,Wineland2001}.

Despite its anticipated high accuracy, the stability of the single-ion based optical clock is severely limited by QPN; long averaging times are required to meet its ultimate accuracy \cite{Wineland1987}.
The measure of the fractional instability is provided by the Allan variance,
$\sigma_y(\tau)=\frac{1}{Q}\frac{1}{\sqrt{ N\tau/\tau_m}}$.
Assuming the transition line $Q\approx 1.6 \times 10^{14}$ \cite{Wineland2001} and a cycle time of $\tau_m\approx 0.1\,{\rm s}$, $4\times 10^7$ measurement cycles are required for a single quantum absorber ($N=1$) to reach $\sigma_y(\tau)=10^{-18}$, corresponding to a total averaging time $\tau$ of a few months. 
For further increase of the stability, the averaging time increases quadratically and will become inordinately long.

One may think of increasing the number of quantum absorbers $N$ as employed in neutral atom based optical standards \cite{PTB2002,Hollberg2001,Ertmer1998}. In this case, however, the atom-laser interaction time sets an upper bound for the $Q$-factor since an atom cloud in free space expands with finite velocity and is strongly accelerated by the gravity during the measurement.
Hence the highest line $Q\approx10^{12}$ \cite{Ertmer1998} obtained for neutral atoms is 2 orders of magnitude smaller than that of a trapped ion.
Furthermore, it has been pointed out that residual Doppler shifts arising from an imperfect wavefront of the probe beam and atom-atom collisions during the measurement affect its ultimate accuracy \cite{PTB2002,Hollberg2001}.

In this Letter, we discuss the feasibility of an ``optical lattice clock" \cite{ref4}, which utilizes millions of neutral atoms separately confined in an optical lattice \cite{lattice1} that is designed to adjust the dipole polarizabilities $\alpha_{E1}$ for the probed electronic states in order to cancel light field perturbations on the measured spectrum \cite{JPSJ}.
In striking contrast with conventional approaches toward frequency standards \cite{Luiten}, the proposed scheme interrogates atoms while they are strongly perturbed by an external field.
We will show that this perturbation can be canceled out to below $10^{-17}$ by carefully designing the light shift potentials.
This scheme permits an exceptionally low instability of $\sigma_y(\tau)\approx10^{-18}$  with an interrogation time of only $\tau=1\,{\rm s}$, which may open up new applications of ultra precise metrology, such as the search for the time variation of fundamental constants \cite{fluct} and the real time monitoring of the gravitational frequency shift.

\begin{figure}
\includegraphics[width=0.9\linewidth]{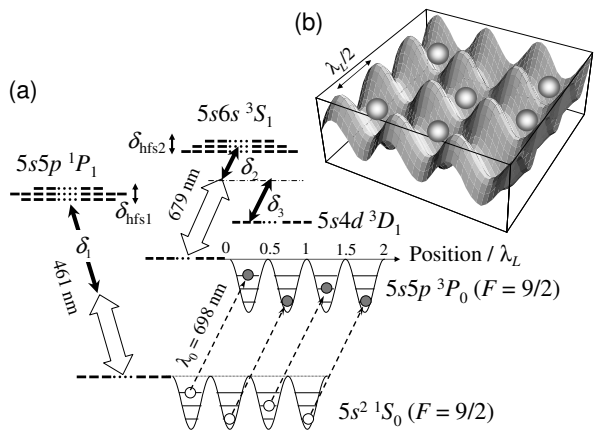}
\caption{ Simplified optical coupling scheme for $^{87}{\rm Sr}$. (a) In the limit of large detunings $\delta_i$ of the coupling laser compared to the hyperfine splittings $\delta_{\rm hfs}$, the squared transition dipole moment of the upper $J$ manifold can be simply added up, resulting in a quasi-scalar light-shift. 
(b) 3D optical lattice provides Lamb-Dicke confinement while it prevents atom-atom interactions.
}
\label{level}
\end{figure}

Figure~\ref{level} illustrates the proposed scheme.
Sub-wavelength confinement provided by the optical lattice localizes atoms in the Lamb-Dicke regime (LDR) \cite{Dicke}, where the first order Doppler shift as well as the photon recoil shift disappears \cite{Bergquist1987,Ido2003} and the second order Doppler shift can be made negligibly small by sideband-cooling atoms down to the vibrational ground state \cite{Wineland1987,Ido2003}.   
In addition, a 3-dimensional lattice with less than unity occupation could
reduce the collisional frequency shifts \cite{PTB2002,collision}.
Therefore this scheme simulates a system where millions of single-ion clocks operate simultaneously.

The transition frequency $\nu$ of atoms exposed to the lattice electric field of $\mathcal{E}$ is described as, 
\begin{equation}\label{l1}
  \hbar\nu = \hbar \nu^{(0)}-\frac{1}{4}\Delta \alpha ({\bf
    e},\omega )\mathcal{E}^2-\frac{1}{64}\Delta \gamma ({\bf
    e},\omega)\mathcal{E}^4-\dots,
\end{equation}
where $\nu ^{(0)}$ is the transition frequency between the unperturbed
atomic states, $\Delta \alpha ({\bf e},\omega)$ and $\Delta \gamma
({\bf e},\omega)$ are the difference between the ac
polarizabilities and hyperpolarizabilities of the upper and lower
states, which in the general case depends both on the light wave
frequency $\omega$ and on the unit polarization vector ${\bf e}$.
Higher order corrections are included in the hyperpolarizability $\gamma ({\bf e},\omega)$ and in the higher-order multipole corrections to the polarizability (magnetic dipole $M1$ and electric quadrupole $E2$ terms in
addition to the electric dipole polarizability $\alpha_{E1}$):
\begin{equation}\label{l1a}
  \alpha ({\bf e},\omega )=\alpha_{E1} ({\bf e},\omega )+\alpha_{M1} ({\bf e},\omega )
  +\alpha_{E2} ({\bf e},\omega ).
\end{equation}
By canceling out the  polarizabilities of the upper and lower states to set $\Delta \alpha ({\bf e},\omega )=0$, the observed atomic transition frequency will be equal to the unperturbed transition frequency independent of the laser intensity $I\propto|\mathcal{E}|^2$, as long as  higher order corrections $O(\mathcal{E}^4)$ are negligible.

We first discuss the cancellation of the electric dipole polarizabilities $\alpha_{E1}({\bf e}, \omega)$ that dominate the light shift perturbations.
In order to provide complete control over perturbations, parameters are preferably specified by the frequency $\omega$ that is the most accurately measurable value in experimental physics.
Other parameters such as the light polarization ${\bf e}$ should have less influence on the system: Our strategy is to employ the $J=0$ state that exhibits a scalar light shift \cite{ref4}.
Here we propose the $5s^2 \,{}^1\!S_0 \rightarrow 5s5p \,{}^3\!P_0$ forbidden transition ($\lambda_0=698\,{\rm nm}$) of $^{87}{\rm Sr}$ with  nuclear spin  $I=9/2$ as the ``clock" transition \cite{ref4}, in which we take advantage of the hyperfine mixing of the $^3\!P_0\,(F=9/2)$ state with the $^{1,3}\!P_1$ states \cite{hfs} to gain a finite lifetime of $\Gamma_0^{-1}=150\,{\rm s}$.
This transition rate turned out to be ideal to perform ultra narrow spectroscopy, as the blackbody radiation at 300~K quenches the $^3\!P$ metastable states of strontium  and gives an effective lifetime of $\sim100\,{\rm s}$ \cite{Yasuda2003}.
Experimentally, however, the attainable line $Q$-factor will be limited by the collision-limited lifetime of 10 s at the vacuum pressure of $10^{-10}\,{\rm torr}$ and/or by the mean photon-scattering time ($\sim 10$ s) of atoms in a far-off-resonant light trap.

\begin{figure}
\includegraphics[width=0.85\linewidth]{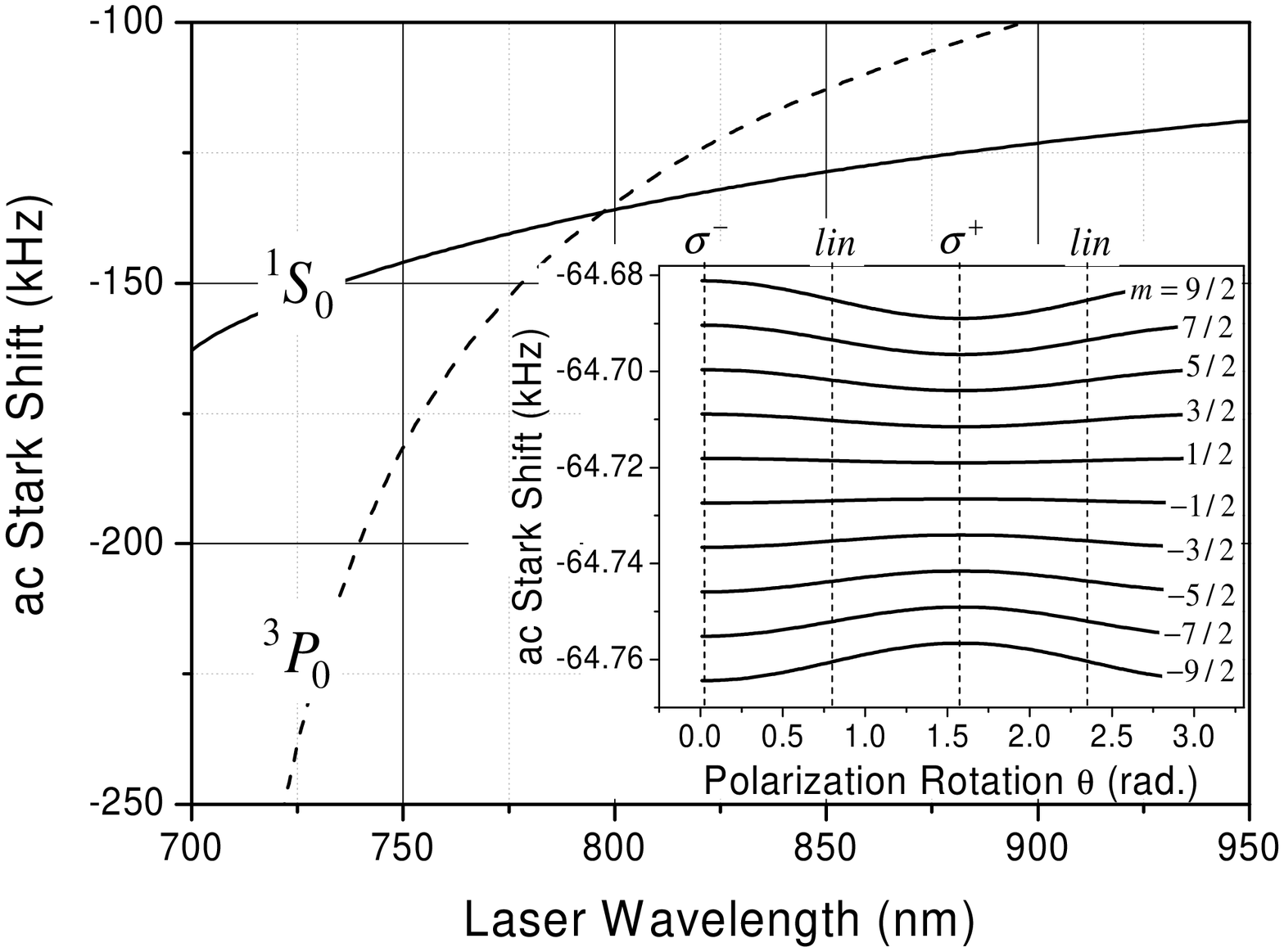}
\caption{ Light shift as a function of the trapping laser wavelength for a laser intensity of $I=10\,{\rm kW}/{\rm cm}^2$. 
The solid and dashed lines show the light shifts for the $^1\!S_0$ and $^3\!P_0$ states, respectively, which intersect at $\lambda_L\approx 800 \, {\rm nm}$.
The inset illustrates the insignificance of the polarization-dependent light shifts of the $^3\!P_0\,(F=9/2)$ state by taking into account the dipole coupling to the $^3\!S_1$ and $^3\!D_1$ state at $\lambda_L$ in presence of a magnetic field $B_0=30\,{\rm mG}$.
}
\label{stark}
\end{figure}

Figure \ref{stark} shows the light shift for the $^1\!S_0$ and  $^3\!P_0$ states as a function of the trapping laser wavelength with an intensity of $I=10\,{\rm kW/cm^2}$.
The calculation is performed by summing up the light-shift contributions with electronic states up to $n=11$ orbits \cite{JPSJ}, in which we employed new values of the dipole moments determined in recent experiments \cite{Ido2003} to find the intersection wavelength at $\lambda_L \approx 800\,{\rm nm}$.
The light shifts around the intersection are mainly determined by the states indicated in Fig.~\ref{level}(a): 
The light shift of the $^3\!P_0$ state can be tuned arbitrarily in the near infrared range, being ``pushed downward" by $^3\!S_1$ and ``pulled upward" by the $^3\!D_1$ state, while that of the $^1\!S_0$ state monotonically decreases toward the dc Stark shift. This tuning mechanism, therefore, can be similarly applied to heavier alkaline-earth (like) atoms. 
At $\lambda_L$ the light shift $\nu_{ac}$ changes with the trapping laser frequency $\omega$ as ${\rm d} \nu_{ac}/{\rm d} \omega=-3.6\times 10^{-10}$ and $-1.3\times 10^{-9}$ for the $^1\!S_0$ and $^3\!P_0$ state, respectively.
This precision enhancement of more than a factor of $10^9$ allows to control the light shift well within 1 mHz by defining the coupling laser within 1 MHz or $10^{-8}$ precision for the optical frequency, which can be easily accomplished by conventional optical frequency synthesis techniques.
Using the same set of dipole moments as used in the above calculation, the blackbody frequency shift \cite{Wing} in the clock transition is estimated to be $-2.4$ Hz at $T=293$ K: The uncertainty can be reduced to the 10 mHz level by controlling the surrounding temperature variation $\Delta T\leq 0.5$ K. Alternatively, by operating the clock at $T=77$ K, the blackbody shift is dramatically decreased to 10 mHz because of its $T^4$ dependence.

Owing to the hyperfine interaction (HFI), the clock transition no longer consists of simple scalar states; therefore the tensor light shift arises.
The fraction of the tensor shift in the total light shift is roughly proportional to $\delta_{\rm hfs}/\delta$, i.e., the ratio of the hyperfine splitting of the coupled electronic state $\delta_{\rm hfs}$ to the trapping laser detuning $\delta$ to that state.
For the $^3\!S_1$, $^3\!D_1$, and $^1\!P_1$ state, as shown in Fig.~\ref{level}(a), the ratio $\delta_{\rm hfs}/\delta$ is $8.1\times 10^{-5}$, $5.4\times 10^{-6}$, and $2.1\times 10^{-7}$, respectively. 
Therefore the tensor shift of the $^3\!P_0$ state can be 2 orders of magnitude larger than that of the $^1\!S_0$ state. 
In order to give an approximate estimate for this tensor shift, we calculated the polarization-dependent light shift in the $^3\!P_0$ state, taking into account the electric-dipole coupling to the $^3\!S_1$ and $^3\!D_1$ hyperfine manifolds that are energy shifted due to HFI.
Both of these manifolds provide half ($\approx 65$ kHz) of the total light shift but dominate its tensor component: Although higher lying electronic states support another 50 \% of the light shift, their contribution to the tensor shift can be less than 50 \%, as the ratio $\delta_{\rm hfs}/\delta$ monotonically decreases for larger $\delta$. 
The inset of Fig.~\ref{stark} shows the result with the light polarization parameterized as
${\bf e}=\cos \theta\,{\bf e}_- +i \sin \theta\,{\bf e}_+$ (${\bf e}_\pm$ represent the unit vector for $\sigma^\pm$ polarization), where a bias magnetic field of $B_0=30\,{\rm  mG}$ is applied to reduce the Raman coherences among the Zeeman substates.
The $m=\pm1/2$ state can be best used for the ``clock" transition, as it exhibits the smallest polarization-dependence of less than 1 Hz.
By employing these states, one could control the light shift within 1 mHz by defining $\theta$ within 1~mrad, even if one applied the linearly polarized trapping laser ($\theta=\pi/4$) where the gradient ${\rm d}\nu_{\rm ac}/{\rm d}\theta\,(=0.83\,{\rm Hz/rad})$ is the largest.

In order to provide an estimate for the higher-order field contributions $O(\mathcal{E}^4)$ described by ac hyperpolarizabilities, we have used the general theory \cite{ref5,ref1,ref2,ref3,refx} for calculating
the light shift for the $5s^2 \,{}^1\!S_0 \rightarrow 5s5p
\,{}^3\!P_0 $ transition on the basis of the
Green's function method in the Fues' model potential approximation \cite{ref1}.
These model potential calculations have reproduced Fig.~\ref{stark}
to within 3 \% accuracy both for the amount of the ac Stark shifts as well as the intersection frequency $\omega_L(=2\pi c/\lambda_L)$, confirming the validity of these two independent approaches.

To calculate the $M1$ and $E2$ contributions to the polarizability Eq.~(\ref{l1a}), the magnetic dipole and electric quadrupole
atom-field interactions should be taken into account together with
the electric dipole term in the amplitude,
\begin{equation}
 \label{l1c}
\hat V({\bf r})=\hat V_{E1}+\hat V_{M1}+\hat V_{E2},
\end{equation}
of the interaction Hamiltonian $\hat H(\mathbf r, t)=\hat V(\mathbf r)e^{-i\omega t} +\hat V^\dag(\mathbf r)e^{i\omega t}$ \cite{ref1}.
The magnetic dipole polarizability for the ground state equals
zero, while for the excited state $5s5p\,^3\!P_0$ its value is
proportional to the squared fine-structure constant ($\alpha=1/137$), to the splitting of the triplet states $E_{10}\equiv E_{5p\,
^3\!P_1}-E_{5p\, ^3\!P_0}$ and to the square of the wave functions
overlapping integral. 
The quadrupole polarizabilities of 
both levels are on the order of $(\alpha\omega)^2$ and may also
become considerable only in the closest vicinity of the resonance on the
quadrupole-allowed transition. Numerical estimates with the
frequency $\omega_L$ where $\Delta \alpha_{E1}(\omega_L)= 0$ gives
$\alpha_{M1}\approx\alpha_{E2} \approx 10^{-7}\times \alpha_{E1} $
for  both levels.

The hyperpolarizability $\gamma ({\bf e},\omega)$ is calculated
starting from a formal expression for the fourth-order
quasi-energy in terms of the field-free wave functions
$|0\rangle$ and the reduced quasi-energy Green's functions $G$
\cite{ref1, ref2}:
\begin{eqnarray}
\label{l2}
    \Delta E^{(4)}&=&-\frac{\mathcal{E}^4}{64}\gamma ({\bf
    e},\omega)
=-\langle\langle 0|HGHGHGH|0\rangle\rangle\nonumber\\
&+& \langle\langle
0|HGH|0\rangle\rangle\langle\langle0|HG^2H|0\rangle\rangle,
 \end{eqnarray}
where the double brackets indicate the integration over the spatial
variables and averaging over time.
In Eq.~(\ref{l2}) only the dipole term of Eq.~(\ref{l1c}) is taken into account
in the interaction Hamiltonian $\hat V(\mathbf r)$.

After the time averaging and angular integration,
the dipole dynamic
polarizability and hyperpolarizability tensors may in general be
resolved into 3 and 5 irreducible parts, $\alpha_p$~($p=0,1,2$) and $\gamma_q$~($q=0,1,2,3,4$), respectively, of which only
scalar terms $\alpha_0$ and $\gamma_0$ contribute in a state with
the total momentum $J=0$ \cite{ref2}. $\alpha_p$ and $\gamma_q$
are determined by linear combinations of frequency-dependent
radial matrix elements of the second, third and fourth orders.

It is to note first that the scalar parts of $\gamma_0 ({\bf
e},\omega)$ are different for the linear and circular type of
polarization, $\gamma_0^l(\omega)\neq \gamma_0^c(\omega)$, even
for atoms in $S$-states, while all the terms of the
polarizability in Eq.~(\ref{l1a}) are independent of ${\bf e}$ for a
state with $J=0$. Secondly, the number and type of singularities
for the hyperpolarizability exceeds that for the polarizability, and
the contribution of these singularities also depend on the
polarization of the light field. E.g., for  linearly polarized
radiation, there are two-photon singularities of
$\gamma_0^l(\omega)$ on the $J=0$ states, while for the
circular polarization such singularities cannot appear.  So, the
hyperpolarizability of the ground state with two equivalent
electrons $5s^2\, ^1\!S_0$ in Sr may be written as
 \begin{eqnarray}
 \label{l2a}
\gamma^l(\omega)&=&\gamma^c(\omega)+\frac89\left[\sigma_{101}(\omega,2\omega,\omega)
+\frac35\Sigma_{121}\right.\nonumber\\
 &-&\left.\frac25\sigma_{121}(\omega,2\omega,\omega)\right]
 \nonumber \\
\gamma^c(\omega)&=&\frac89\left[\Sigma_{101} +\frac15\Sigma_{121}
+\frac65\sigma_{121}(\omega,2\omega,\omega)\right]\nonumber\\
&-&2\alpha_0(\omega)S_{-3}(\omega)
\end{eqnarray}
where $\alpha_0(\omega)$
is the polarizability and $S_{-3}$ is the so-called frequency-dependent oscillator strengths moment.
The following notations for the radial matrix elements and their
combinations were used above:
\begin{eqnarray}
 \label{l2d}
\Sigma_{l_1l_2l_3}&=&\sigma_{l_1l_2l_3}(\omega,0,\omega)+
\sigma_{l_1l_2l_3}(\omega,0,-\omega);
 \nonumber \\
\sigma_{l_1l_2l_3}(\omega_1,\omega_2,\omega_3)&=&R_{l_1l_2l_3}(\omega_1,\omega_2,\omega_3)\nonumber\\
&+&R_{l_1l_2l_3}(-\omega_1,-\omega_2,-\omega_3); \nonumber \\
R_{l_1l_2l_3}(\omega_1,\omega_2,\omega_3)&=&\langle
0|rg_{l_1}^{\omega_1}rg_{l_2}^{\omega_2}rg_{l_3}^{\omega_3}r|0\rangle.
\end{eqnarray}
The arguments in Eqs.~(6) indicate the frequency dependence in energy denominators \cite{ref1}. Here, $g_l^\omega $ is the radial Green's function in the subspace of the jumping electron's states with angular momentum $l$.

In the vicinity of a one-photon resonance with frequency
detuning $|\delta| \ll \omega$, the product of the second-order and
third-order matrix elements in Eq.~(\ref{l2}), which corresponds to the term $\alpha_0S_{-3}$ in Eq.~(\ref{l2a}), dominates the third-order
poles of order $\delta^{-3}$. The fourth-order matrix element in Eq.~(\ref{l2}) has the second-order poles ($\delta^{-2}$).
The two-photon ($2\omega$) resonance
singularity on the $J=0$ states in the radial matrix elements
$R_{101}(\omega, 2\omega,\omega)$ appears only for
$\gamma^l(\omega)$ and that on the $J=2$ states in
$R_{121}(\omega,2\omega,\omega)$ appears both for $\gamma^l(\omega)$ and
$\gamma^c(\omega)$.

We used the analytic Sturm-series representation of the Green's
function, corresponding to
calculation of the infinite sums over the total atomic spectrum
including the continuum.
Finally, the radial integrals in Eq.~(\ref{l2d}) were presented in the
form of absolutely converging series, well suited for the
numerical computations.
The numerical results for $\gamma_{^1\!S_0}^l (\omega)$ and
$\gamma_{^3\!P_0}^l (\omega)$ at the intersection frequency of $\omega_L$
are $6.3\cdot10^6$ a.u. and $2.7\cdot10^8$ a.u., respectively, which give  fourth-order ac Stark shifts of 
$\Delta E_{^1\!S_0}^{(4)}/h\approx -5.3\cdot 10^{-5}~{\rm Hz}$ and 
$\Delta E_{^3\!P_0}^{(4)}/h\approx -2.3\cdot 10^{-3}~{\rm Hz}$ for the trapping laser intensity of $10\,{\rm
kW/cm}^2$.
Therefore the contribution of the higher order light shifts is as small as $5\times 10^{-18}$.
Further elimination of this systematic error can be achieved by extrapolating the trapping laser intensity to zero \cite{Ido2003} in a quadratic way. 



In summary, we have discussed the feasibility of precision spectroscopy of neutral atom ensembles confined in an optical lattice, by applying a light-shift cancellation technique on the $^1S_0\,(F=9/2,m_F=\pm1/2)\rightarrow{}^3P_0\,(F=9/2,m_F=\pm1/2)$ clock transition of $^{87}{\rm Sr}$ that has a negligibly small tensor shift and a suitable transition moment to perform spectroscopy.
Our calculation including the higher-order Stark shift confirmed that the measurement of the unperturbed atomic transition at 1 mHz level is feasible, allowing for the projected accuracy of better than $10^{-17}$.
Since this scheme is equivalent to  millions of single-ion-clocks operated in parallel, excellent improvement of the stability $\sim \sqrt {10^6}$ can be expected.
The theory and method for the numerical calculation of the
hyperpolarizability tensor for arbitrary alkaline-earth atoms will be discussed in a forthcoming paper.

H. K. acknowledges financial support from Japan Society for the Promotion of Science under Grant-in-Aid for Young Scientists (A) KAKENHI 14702013, and T. Kurosu and T. Ido for useful conversation.

Note Added: After submission of the paper, we have experimentally demonstrated the optical lattice clock and determined the intersection wavelength \cite{Takamoto2003}.


\begin{thebibliography}{0}
\bibitem{collision} Assuming the proposed clock transition at  $\lambdabar=698\,{\rm nm}/2\pi$ with linewidth $\Gamma_0$, the resonant dipole-dipole frequency shift $\Gamma_0 (\lambdabar/d)^3$ is on the order of $10^{-5}\,{\rm Hz}$ for atoms separated by the lattice period of $d=400\ {\rm nm}$.
\end{thebibliography}
\end{document}